\documentclass[]{spie}  

\usepackage{amsmath,amsfonts,amssymb}
\usepackage{graphicx}
\usepackage[bookmarks=false,colorlinks=true, allcolors=blue]{hyperref}

\title{Performance and early science with the Subaru Coronagraphic Extreme Adaptive Optics project}

\author[a,b]{Thayne Currie}
\author[b]{Olivier Guyon}
\author[b]{Julien Lozi}
\author[c]{Tyler Groff}
\author[d]{N. Jeremy Kasdin}
\author[e]{Frantz Martinache}
\author[f]{Timothy D Brandt}
\author[g]{Jeffrey Chilcote}
\author[h,i]{Christian Marois}
\author[i]{Benjamin Gerard}
\author[j]{Nemanja Jovanovic}
\author[b]{Sebastien Vievard}
\affil[a]{NASA-Ames Research Center, Moffett Field, California, USA}
\affil[b]{Subaru Telescope, 650 N. Aohoku Pl., Hilo, Hawai'i, USA}
\affil[c]{NASA-Goddard Space Flight Center, Greenbelt, MD, USA}
\affil[d]{Princeton University, Princeton, NJ, USA}
\affil[e]{Univ. CÃŽte d'Azur, Nice, France}
\affil[f]{University of California-Santa Barbara, Santa Barbara, CA, USA}
\affil[g]{University of Notre Dame, South Bend, IN, USA}
\affil[h]{NRC-Herzberg, Victoria, BC, Canada}
\affil[i]{University of Victoria, BC, Canada}
\affil[j]{California Institute of Technology, Pasadena, CA, USA}

\authorinfo{Further author information: (Send correspondence to T. Currie)\\T. Currie: E-mail: currie@naoj.org}

\begin{document} 
\maketitle

\begin{abstract}
We describe the current performance of the Subaru Coronagraphic Extreme Adaptive Optics (SCExAO) instrument on the Subaru telescope on Maunakea, Hawaii and present early science results for SCExAO coupled with the CHARIS integral field spectrograph.   SCExAO now delivers H band Strehl ratios up to $\sim$ 0.9 or better, extreme AO corrections for optically faint stars, and planet-to-star contrasts rivaling that of GPI and SPHERE.  CHARIS yield high signal-to-noise detections and 1.1--2.4 $\mu m$ spectra of benchmark directly-imaged companions like HR 8799 cde and kappa And b that clarify their atmospheric properties. We also show how recently published as well as unpublished observations of LkCa 15 lead to a re-evaluation of its claimed protoplanets.    Finally, we briefly describe plans for a SCExAO-focused direct imaging campaign to directly image and characterize young exoplanets, planet-forming disks, and (later) mature planets in reflected light.
\end{abstract}


\section{INTRODUCTION}
\label{sec:intro}  
Starting about a decade ago, ground-based telescopes using facility (conventional) adaptive optics (AO) systems revealed the first direct images of young, self-luminous superjovian mass planets orbiting nearby stars  \cite{Marois2008,Marois2010,Lagrange2010,Rameau2013,Currie2014}.  Follow-up multi-wavelength photometry provided the first constraints on their atmospheric properties, revealing the planets to be redder and cloudier/dustier than substellar objects with the same temperatures and, in some cases, showing evidence for non-equilibrium carbon chemistry\cite{Currie2011,Galicher2011}.   Spectra for the first directly-imaged planets revealed evidence for low surface gravities and the presence of multiple molecular species \cite{Barman2011,Currie2014}.    Near-infrared (IR) conventional AO systems achieve typical planet-to-star contrasts typically of 10$^{-3}$, $10^{-4}$, and 10$^{-5}$ to 5$\times$10$^{-6}$ at angular separations of 0.25", 0.5" and  1.0"\cite{Brandt2014c}.   Most imaged exoplanets discovered with conventional AO lie beyond 0.5".   Surveys with these systems are typically sensitive to only the most massive planets (10--15 $M_{\rm J}$) at orbits exterior to those in our own solar system ($a_{\rm p}$ $\gtrsim$ 30--100 au)\cite{Brandt2014a}.

Now, \textit{extreme} AO systems like the \textit{Gemini Planet Imager} (GPI) on Gemini-South and \textit{Spectro-Polarimetric High-contrast Exoplanet REsearch instrument} (SPHERE) on the Very Large Telescope (VLT) are able to detect planets at 0.25"--1.0" a factor of 10-100 times fainter in the near-IR \cite{Macintosh2014,Vigan2015}.  These deeper contrasts have opened up new exoplanet discovery space, probing planets that are lower mass (down to $\sim$ 2 $M_{\rm J}$) and closer in angular separation ($\rho$ $\sim$ 0.1"--0.4") and physical projected separation ($a_{\rm p}$ $\sim$ 10-15 au) \cite{Macintosh2015,Chauvin2017,Currie2015,Milli2017,Keppler2018}.   Integral field spectrographs coupled to GPI and SPHERE have provided new insights into the atmospheric properties of young, jovian planets, including their cloud cover, temperature, and gravities (e.g. \cite{Bonnefoy2016,Chilcote2017,Rajan2017}).   Extreme AO surveys have provided new constraints on the frequency of 5--13 $M_{\rm J}$ planets at 10--30 au, suggesting that these companions form a separate population from more massive brown dwarf companions\cite{Nielsen2019}.

Despite recent technological advances utilized by these extreme AO systems, only about 20 or so exoplanets have been directly imaged thus far.   Direct imaging preferentially detects luminous (massive), wide-separation planets.  However, analyses of radial-velocity searches for mature planets suggest that the frequency of planets at the innermost separations probed by the latest extreme AO surveys is small compared to a peak at $\sim$ 2--3 au \cite{Fernandes2019}.   Even at wider separations (10--30 au), typical sensitivities are on the order of $\sim$ 3--5 $M_{\rm J}$ \cite{Nielsen2019}.    Yet the frequency of gas giant planets comparable to or lower than Jupiter's mass is significantly higher than that of superjovian planets \cite{Fernandes2019}.  The yield of direct imaging surveys appears to be biased towards stars more massive than the Sun.   Directly detecting planets on smaller orbits, with lower masses, and around a wider range of stellar masses requires significantly better performance at small angular separations, a higher-fidelity AO correction at wider separations, and better performance for optically fainter stars.


  Here, we provide an update on the performance and early science obtained with the Subaru Coronagraphic Extreme Adaptive Optics project (SCExAO) at the Subaru Telescope on Maunakea \cite{Jovanovic2015}.   SCExAO employs improvements in wavefront control hardware and software improving its performance at small angular separations and for faint guide stars.   It will mature key advances in wavefront control and coronagraphy needed to move beyond recent exoplanet direct imaging capabilities to those capable of imaging an Earth with ELTs. 

\begin{figure}[ht]
   \begin{center}
  \centering
   \includegraphics[scale=0.4,clip]{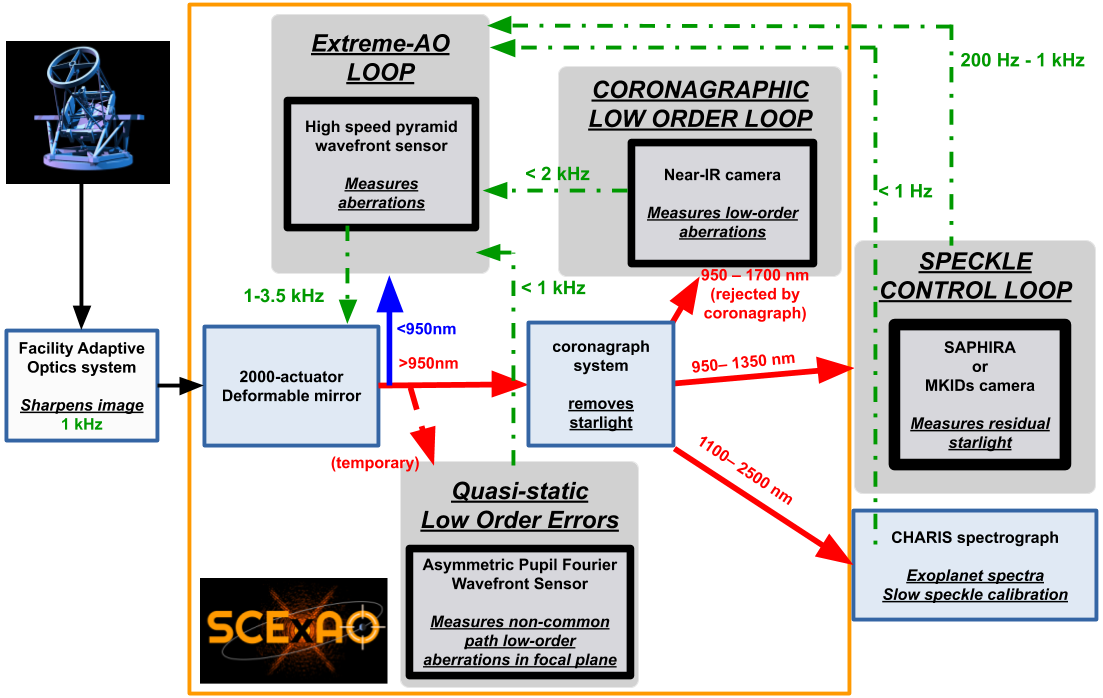}
   \end{center}
   \caption
   { \label{fig:wfcloops} 
       Current schematic of SCExAO.  Note that the coronagraphic low-order loop is not in normal operation; the MKIDs camera (MEC) is undergoing commissioning.
}
\end{figure}
\section{SCExAO Design and Performance}


The standard mode of SCExAO science operation is shown in Figure \ref{fig:wfcloops}.   The system takes in and further sharpens partially-corrected light from Subaru's facility AO system (A0-188) as input, which typically achieves $\sim$ 30--40\% Strehl at 1.6 $\mu m$.   SCExAOâs main wavefront control loop includes a 2,000-element MEMS deformable mirror from Boston MicroMachines driven by a modulated Pyramid wavefront sensor using an OCAM$^{2}$K camera from First-Light Imaging operating over a 600--900 nm bandpass.   The loop can run at speeds up to 3.5 kHz.  However, currently in normal science operations it operates at 2 kHz for bright stars coupled with predictive wavefront control (see below) to decrease wavefront sensor noise error.   For very faint stars($I$ $>$ 9), we typically run the loop at 1 kHz.   The loop can correct for up to 1200 modes of dynamic aberrations.

\begin{figure}[ht]
   \begin{center}
  \centering
   \includegraphics[scale=0.65,clip]{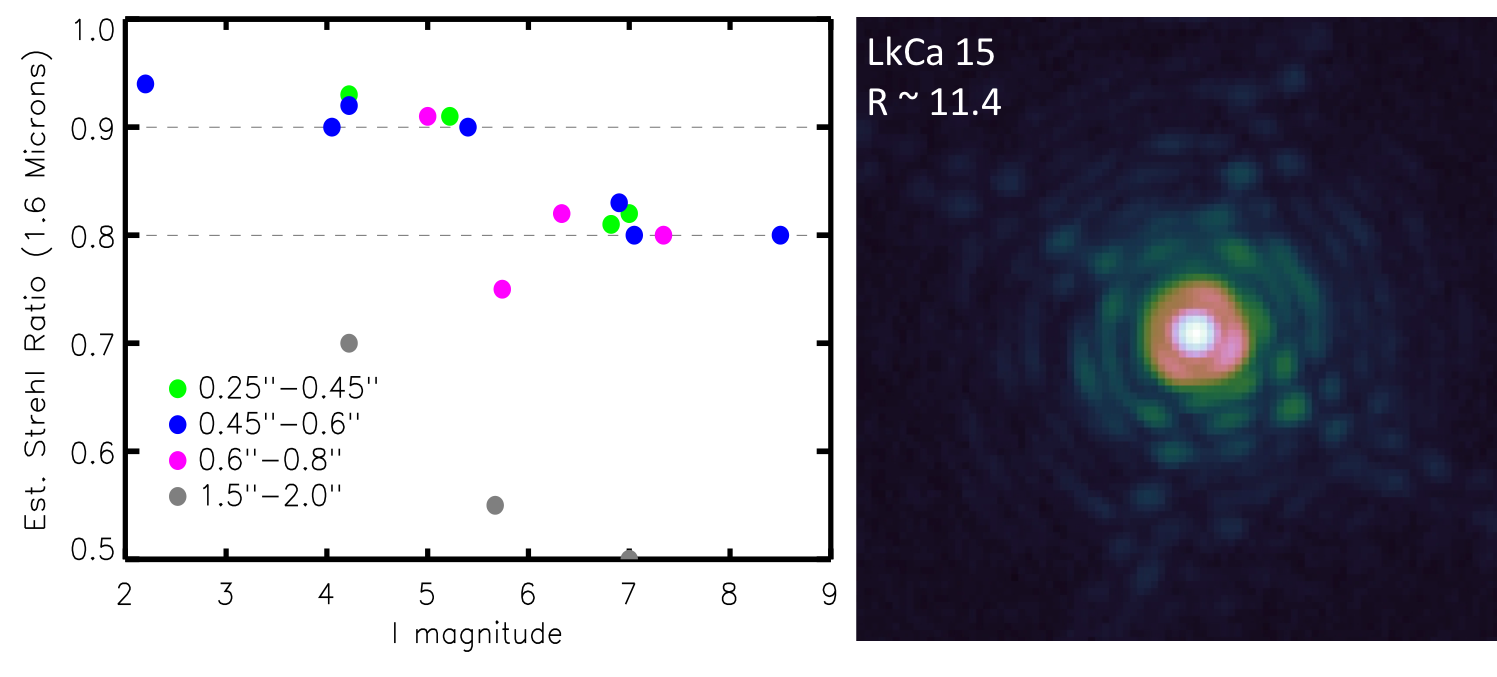}
   \end{center}
   \caption
   { \label{fig:strehl} 
    (left) Plot of Strehl ratio -- as estimated from the residual wavefront error -- vs. guide star $I$ band magnitude.  (right) SCExAO PSF for LkCa 15 at $K$ band, an optically faint star revealing numerous Airy rings.   
}
\end{figure}

Figure \ref{fig:strehl} presents the AO performance of SCExAO as a function of atmospheric conditions and $I$-band magnitude.   For extremely bright stars ($I$ = 1--5), SCExAO can deliver excellent AO corrections, with estimated Strehl ratios (at 1.6 $\mu m$) of 0.9-0.94 for slightly above average to top-quartile seeing conditions on Maunakea (left panel).     Estimated Strehl ratios drop to $\sim$ 0.75-0.8 for much fainter stars ($I$ $\sim$ 7-8), or below for below-average conditions, although in the former case some of the estimated drop may be due to noise in the estimate of the wavefront correction residuals.   Under the best conditions ($\theta$ $\sim$ 0.2"-0.5"), SCExAO can deliver sharp AO corrections for stars as faint as R $\sim$ 11.4--12, such as LkCa 15 or V819 Tau (right panel).    

Corrected starlight is fed into different optical/near-infrared (IR) science instruments.   For most exoplanet imaging programs, the near-IR integral field spectrograph CHARIS is used.   CHARIS operates in two modes.   The primary workhorse configuration is its low resolution (R $\sim$ 20)  âbroadband modeâ covering the three major near-IR bandpasses (J, H, and K) simultaneously, which makes the instrument well-suited for exoplanet discovery and coarse spectral characterization.   A higher-resolution (R $\sim$ 70) mode in J, H, or K allows for more detailed, follow-up characterization.   For broadband mode, we typically use a Lyot coronagraph with a $\sim$ 0.1" occulting spot; for $H$ band the vector vortex coronagraph is used.   Satellite spots placed at 15.5 $\lambda$/D are used for spectrophotometric calibration and image registration.   

The CHARIS Data Reduction Pipeline converts raw data into data cubes\cite{Brandt2017}.   There is currently no standardized, public pipeline for subsequent data reduction steps (e.g. image registration, PSF subtraction).   However, T.C. has developed data reduction tools for these steps that will likely separately become part of a supported pipeline component\cite{Currie2018a,Currie2018b}.  Other collaborators have successfully carried out at least many data reduction steps using the public PyKLIP package\cite{Wang2015} or their own proprietary codes\cite{Marois2014,Gerard2019}.


\begin{figure}[ht]
   \begin{center}
  \centering
   \includegraphics[scale=0.35,clip]{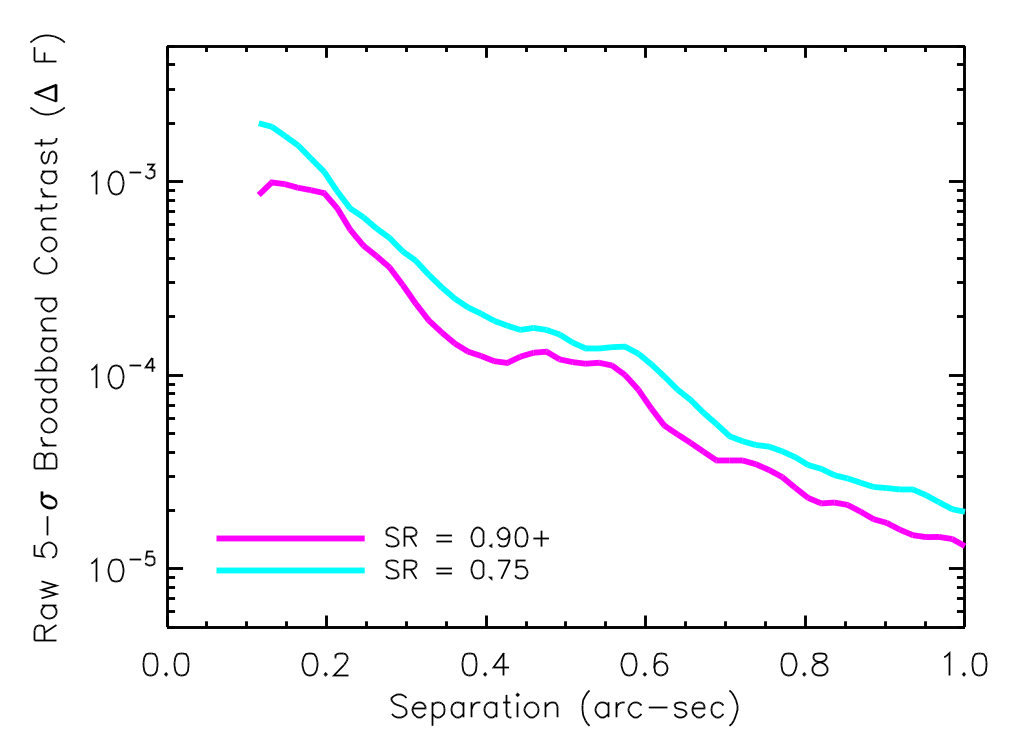}
   \end{center}
   \caption
   { \label{fig:rawcontrast} 
   Raw 5-$\sigma$ broadband contrast obtained for stars at S.R. $\sim$ 0.75 and $\sim$ 0.9 (corrected for finite sample sizes) but without fully masking residual light from satellite spots. Towards the highest Strehl ratios, the raw contrast curve is nearly flat from 0.3"--0.6". The curves are computed by measuring, at each radius, the surface brightness standard deviation (assumed here to be the noise), and multiplying it by 5.
}
\end{figure}

As shown in \ref{fig:rawcontrast}, the raw contrast achieved with SCExAO/CHARIS under very good correction (S.R. $\sim$ 0.9) is roughly flat between $\sim$ 0.3" and 0.6" at $\approx$ 10$^{-4}$, $\sim$ 3$\times$10$^{-4}$ at 0.25", and $\sim$ 2$\times$10$^{-5}$ at 0.8".  At small separations, SPHERE achieves deeper raw contrast\cite{Beuzit2019}, plausible due to its more optimal choice of coronagraphs to suppress halo light compared to our use of a standard Lyot coronagraph, among other possible factors.  

After using advanced post-processing, it achieves contrasts of $\sim$ 10$^{-5}$ at 0.25" and 10$^{-6}$ at 0.5" over 1 hour-long sequences and under good conditions.   These values are comparable to typical contrasts achieved with the Gemini Planet Imager in $H$ band.   Improvements in wavefront control and coronagraphy will likely lead to substantial gains in contrast at small separations.

\begin{figure}[ht]
   \begin{center}
  \centering
   \includegraphics[scale=0.675,clip]{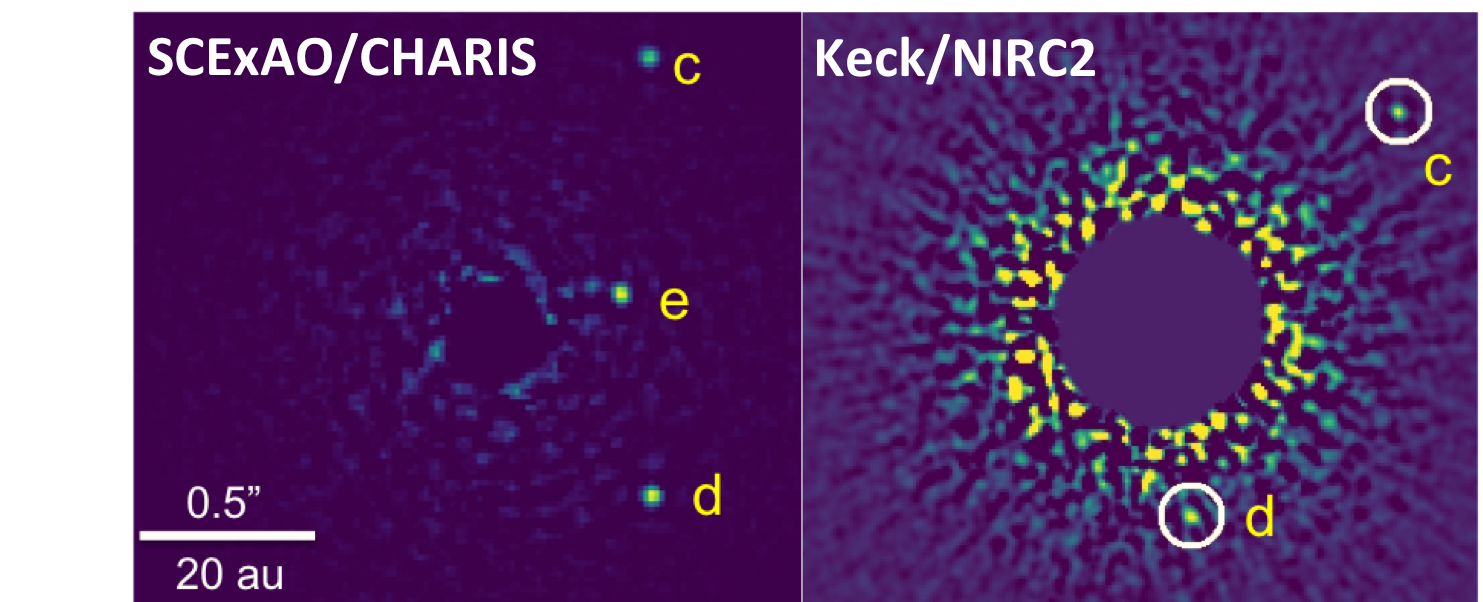}
   \end{center}
   \caption
   { \label{fig:hr8799comp} 
    (left) SCExAO/CHARIS HR 8799 image (wavelength-collapsed) compared to Keck/NIRC2 $H$ band image (right), showing stronger detections of HR 8799 cd and detection of HR 8799 e.   HR 8799 b lies outside CHARIS's field of view.  The CHARIS data were reduced first using advanced PSF subtraction\cite{Currie2012b} to remove the halo in angular differential imaging (ADI) and then simple classical spectral different imaging (a median-combination of channels magnified by wavelength).   The Keck/NIRC2 data were reduced in ADI only.
}
   \end{figure}

Figure \ref{fig:hr8799comp} uses observations of HR 8799 to illustrate the practical performance gain for SCExAO/CHARIS over conventional AO facilties. SCExAO/CHARIS yields a $\sim$ 23-$\sigma$ detection of HR 8799 e -- invisible in the Keck/NIRC2 data -- and much stronger detections of the outer two planets HR 8799 cd.   While spectral differential imaging does result in a contrast gain, a key advantage of SCExAO/CHARIS appears to be the relatively strong temporal correlation of the halo, making post-processing with angular differential imaging (ADI; \cite{Marois2006}) more effective\cite{Gerard2019}.

 \section{Early SCExAO/CHARIS Science Results}
 
 The first year of full science operations for SCExAO and CHARIS has yielded seven peer-reviewed publications\cite{Currie2018a,Currie2018b,Currie2019a,Goebel2018,AsensioTorres2019,Rich2019,Gerard2019} providing new insights about massive exoplanets, planet candidates, protoplanetary and debris disks, brown dwarfs and low-mass stars, and the efficacy of advanced coronagraphy and PSF subtraction methods.   Here, we focus on two systems -- $\kappa$ Andromedae and LkCa 15 -- for which SCExAO/CHARIS data provided new key insights about (proposed) planetary companions.
 
 \subsection{SCExAO Observations of $\kappa$ And}
 
 The directly-imaged low-mass companion to the B9V star $\kappa$ Andromedae ($\kappa$ And b; \cite{Carson2013}) is an object whose properties were better constrained with SCExAO/CHARIS data.  Depending on whether the system $\kappa$ And was a young star with an age (i.e. 30--40 $Myr$) comparable to moving groups like the Columba association or an older, $\sim$ 200 $Myr$-old one, $\kappa$ And b could have a planet-like mass or one consistent with low-mass brown dwarfs\cite{Carson2013,Hinkley2013}.  Interferometric observations of the star favor a young age\cite{Jones2016}.   A spectrum could better determine whether $\kappa$ And b bears a greater resemblance to young, planet-mass brown dwarfs or intermediate-aged more massive brown dwarfs.
 
 \begin{figure}[ht]
   \begin{center}
  \centering
   \includegraphics[scale=0.5,clip]{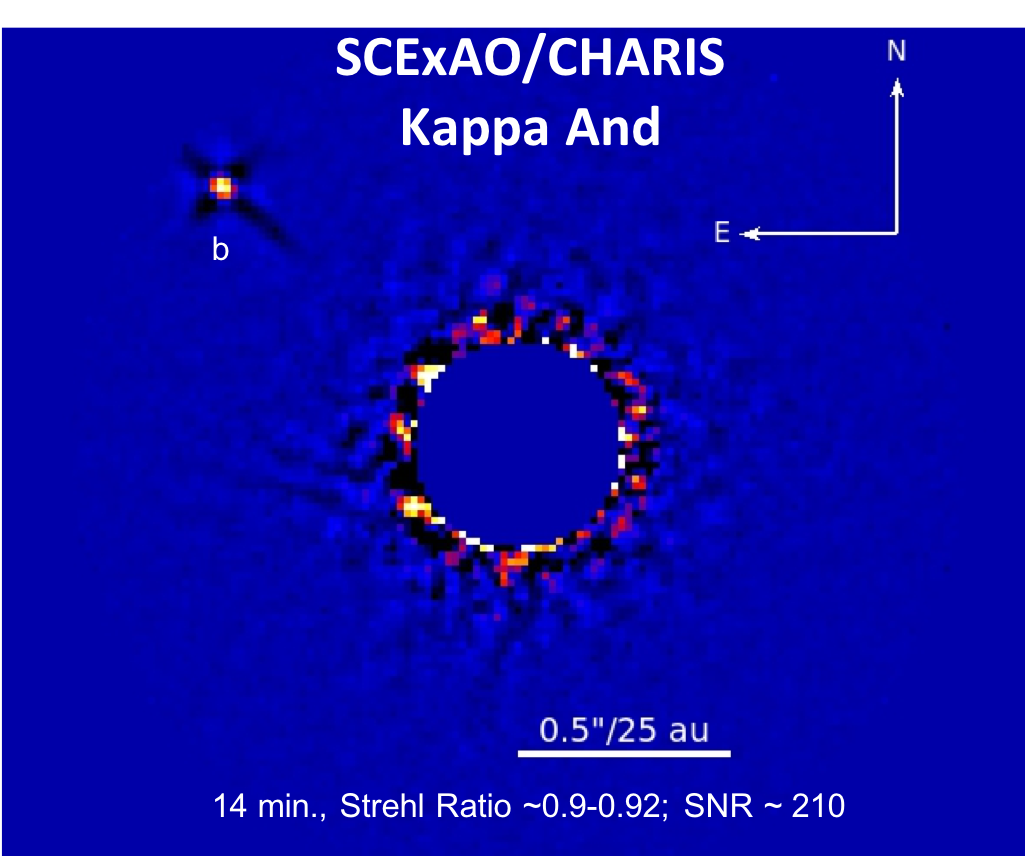}
   \end{center}
   \caption
   { \label{fig:kapandimage} 
    SCExAO/CHARIS image of $\kappa$ And b from \cite{Currie2018a}.
}
   \end{figure}
\begin{figure}[ht]
   \begin{center}
  \centering
   \includegraphics[scale=0.7,clip]{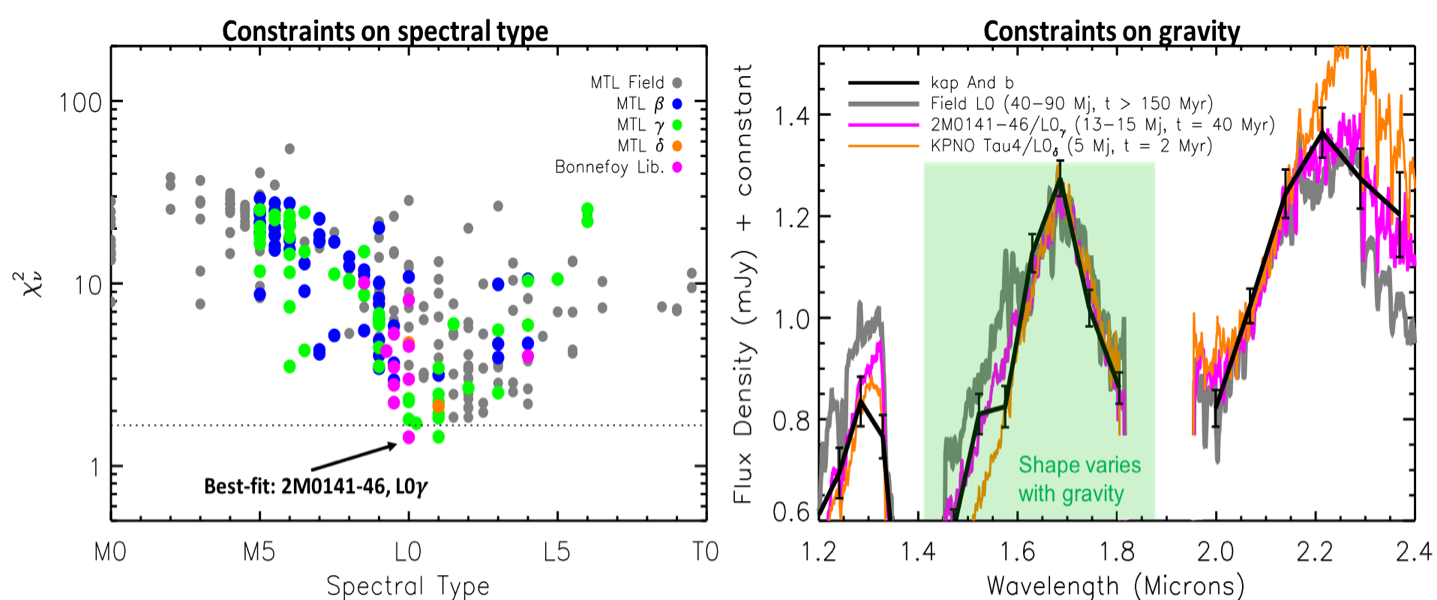}
   \end{center}
   \caption
   { \label{fig:kapandspec} 
    Spectral characterization of $\kappa$ And b with SCExAO/CHARIS, constraining spectral type (left) and gravity (right).   
}
   \end{figure}

 SCExAO/CHARIS data yielded an exceptionally high SNR JHK spectrum of $\kappa$ And b and achieved deep contrast limits down to 0.25" ($\sim$ 12.5 au) despite only 14 minutes of integration time (Figure \ref{fig:kapandimage}).   The shape of $\kappa$ And b's spectrum was best matched by young L0--L1 dwarfs (Figure \ref{fig:kapandspec}).   The $H$-band peak of $\kappa$ And b showed strong evidence for the companion being a low gravity object, consistent with a planetary mass. 
 
 Recent spectral energy distribution modeling of the CHARIS spectrum along with complementary photometry at $Y$ band and in the thermal infrared constrain its temperature to be 1700--1900 $K$ and favor a surface gravity of log(g) $\sim$ 4.0--4.5 (Uyama et al. 2019, submitted).   Orbital modeling of $\kappa$ And b astrometry suggests that the companion has a high eccentricity and a semimajor axis greater than $\sim$ 75 au.   As $\kappa$ And b is at a separation where forming massive planets by core accretion is exceptionally difficult, it likely formed through other means, such as gravitational instability\cite{Boss1997}.
 
 \subsection{SCExAO Observations of LkCa 15}

 The infant Sun-like star LkCa 15 has long been regarded as a key laboratory for studying planet formation.   The star's massive, gaseous protoplanetary disk contains multiple dust components, where a sub-au scale hot disk and cold outer disk are responsible for most of the system's broadband infrared emission \cite{Espaillat2007}.   The spatially resolved cavity separating these two dominant  components\cite{Andrews2011} is a tell-tale sign that much of the disk material has already been incorporated into massive, still-forming ``protoplanetsâ.  
 
 Two studies using sparse aperture masking interferometry and a separate $H_{\rm \alpha}$ imaging identified up to three protoplanets orbiting LkCa 15\cite{Kraus2012,Sallum2015} (LkCa 15 'bcd'). However, recent polarimetric imaging reveal additional dust material at separations comparable to LkCa 15 bcd, whose scattered light signal could possibly be misinterpreted as planets.   SCExAO/CHARIS observations from September 2017 were of sufficient quality to directly detect planets or disk material around LkCa 15, despite the star's faint optical brightness.


    \begin{figure}[ht]
   \begin{center}
  \centering
   \includegraphics[scale=0.7,clip]{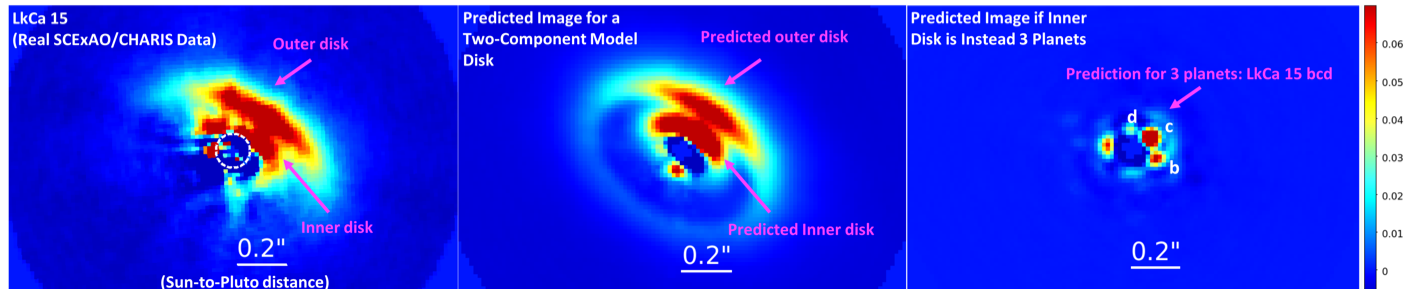}
   \end{center}
   \caption
   { \label{fig:lkca15results} 
   Image of LkCa 15 obtained with SCExAO/CHARIS on September 7, 2017 (left), forward-modeled image for a two-component disk from a theoretical model (middle), and forward-modeled image for three planets derived from previous aperture masking data.   
}
   \end{figure} 
SCExAO/CHARIS data and complementary Keck/NIRC2 data spatially resolved the inner regions of LkCa 15 over same wavelengths where LkCa 15 bcd were originally identified.  However, they showed that emission attributed LkCa 15 bcd predominantly comes from the arc-like forward-scattering peak of the inner disk component previously seen in polarimetry\cite{Currie2019a}(Figure \ref{fig:lkca15results}).   The emission  has the same brightness previously attributed to planets around LkCa 15 in \cite{Sallum2015}.  

Forward-models of LkCa 15 bcd are inconsistent with the SCExAO/CHARIS and Keck/NIRC2 data.  However, scattered light disk models generally reproduced the morphology of emission seen by both instruments.  Complementary data taken with the Keck Observatory helped establish that this arc-like structure is static over time and therefore better consistent with a fixed structure like a disk than orbiting planets.

\begin{figure}[ht]
   \begin{center}
  \centering
   \includegraphics[scale=0.9,clip]{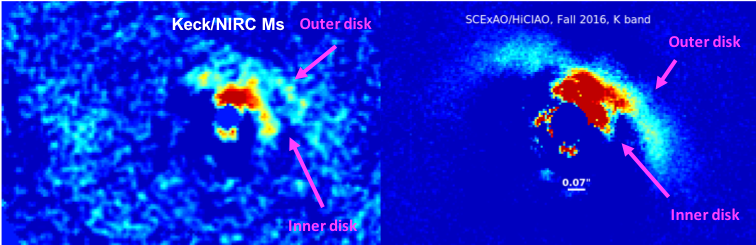}
   \end{center}
   \caption
   { \label{fig:lkca15new} 
   Unpublished SCExAO/HiCIAO and Keck/NIRC2 images of LkCa 15.   Both data sets show evidence for two separate disk components but not directly-imaged planets.}
   \end{figure}

Unpublished data appear consistent with a disk interpretation for most of the emission seen around LkCa 15 at small angles (Figure \ref{fig:lkca15new}).   SCExAO/HiCIAO K-band data obtained in 2016 likewise show a continuous arc of emission.  Using archival Keck/NIRC2 data, we have also successfully resolved the inner and outer disk components at $M_{\rm s}$ ($\lambda_{\rm o}$ $\sim$ 4.67 $\mu m$).   In both data sets, we find no direct evidence for protoplanets around LkCa 15.   

Irrespective of our evidence that emission around LkCa 15 is static, claiming evidence for orbiting protoplanets from aperture masking data is challenging.  Astrometric error bars reported for LkCa 15 `bcd' are large.   It is possible that variable $u--v$ coverage between epochs can induce apparent astrometric offsets when a binary model is assumed in the image reconstruction process (C. Cacares 2019, in prep.).   The choice of calibrators, and whether or not they have circumstellar material, is also important.   As our previous study notes, previous aperture masking data for LkCa 15 is not faithful reproducing the spatial structure of emission at small angles\cite{Currie2019a}.  In other systems with previously claimed protoplanets like HD 169142, structures within the disk could themselves be orbiting\cite{Gratton2019}.   Distinguishing between planet and disk also requires a forward-model of each emission source through the data.

Despite the above cautions, it is still possible that future observations will reveal a set of orbiting protoplanets, perhaps even at the locations previously proposed for LkCa 15 'bcd'.   Such companions, though, would be significantly fainter than previously reported: e.g. at a contrast of $\sim$ 10$^{-3}$ instead of $\sim$ 10$^{-2}$.   Obtaining accurate photometry and astrometry will be challenging given the appearance of such companions against a bright inner disk.

\section{Future Plans for SCExAO}

      \begin{figure} [ht]
   \begin{center}
  \centering
   \includegraphics[scale=1,clip]{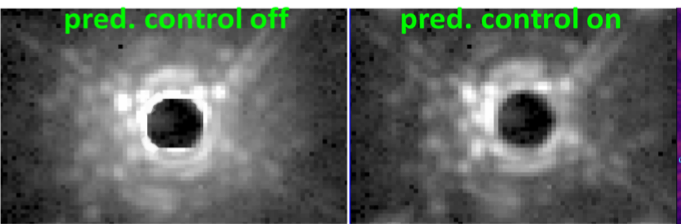}
   \end{center}
   \caption
   { \label{fig:pred}
   On-sky SCExAO PSF without (left) and with (right)  predictive control.   The images are of the same star and are averages of 54 consecutive 0.5s images taken 3 minutes apart (26s total).
}
   \end{figure}

\subsection{Technical Advances}
A detailed overview of recent and upcoming technical improvements to SCExAO is discussed in Lozi et al. \cite{Lozi2018}: a subset of these upgrades are discussed below, particularly focusing on wavefront sensing/control advances and new science modes.   

\begin{itemize}
    \item \textbf{Predictive Wavefront Control} -- We have now successfully implemented ``predictive control" using empirical orthogonal functions\cite{Males2018}.   Predictive control allows us to reduce the servo lag error in the wavefront sensing error budget, improving contrast at small angular separations.   Figure \ref{fig:pred} shows a demonstration of predictive control, where it yields a factor of $\sim$ 3 improvement in contrast at small angular separations.
    
    \item \textbf{CRED-2 Detector} - We have replaced SCExAO's internal short-wave infrared camera (170 Hz, 100 e$^{-}$ read noise) with a new CRED-2 detector from First-Light, which can run in excess of 1 kHz in subframe mode, with low, $<$ 30 e$^{-}$ read noise and utilized for focal-plane wavefront control.  The CRED-2 allows precise alignment of a coronagraph using both focal plane and pupil plane viewing modes.   The CRED-2 can also be used to obtain broadband science images at wavelengths adjacent to those being used by CHARIS.   
    
    \item \textbf{MEC} - A more fundamental advance is MEC, an 20,000 pixel MKIDS-based detector built by the University of California-Santa Barbara.  MEC is a noiseless, ultra-cooled photon-counting detector able to measure the energy and wavelength of every photon.   This capable makes MEC a powerful tool for reducing chromatic effects in the speckle halo and for driving fast, focal-plane wavefront sensing and control (e.g. speckle nulling).
    
    \item \textbf{New Focal-Plane Wavefront Sensing Techniques} -- In addition to speckle nulling, additional focal-plane techniques may be tested to allow SCExAO to achieve a far deeper, sustained dark hole than previously possible.   In particular, in collaboration with NASA-Ames Research Center, the SCExAO team is developing and eventually planning to implement (on sky) \textit{Linear Dark Field Control}, which uses the linear response of perturbations in the bright, uncorrected region to maintain a dark hole originally dug through focal-plane wavefront sensing techniques\cite{Miller2017}.   See SPIE paper 11117-65 for more details\cite{Currie2019}. 
    
    \item\textbf{Integral Field Polarimetry with CHARIS} -- Recently commissioned, CHARIS's ``integral field polarimetry" mode splits light into two polarizations over a 1.2"x2.4" field for 22 separate spectral channels covering J through K.   This new mode allows us to detect small-scale emission from protoplanetary and debris disks and yield both total intensity and polarized intensity spectra to get a wavelength-dependent polarization fraction.    Its utility for planet detection around young stars is key.   Specifically, it can allow us to measure both total and polarized intensity will help us to determine whether candidate point source-like features in disks are highly polarized. Having total intensity IFS data for any point sources detected will also be critical to help characterize these sources, to assess whether their spectra look like substellar objects or CPD-dominated objects.
    
\end{itemize}

On a slightly longer timescale,  the key advance will be to replace the facility AO system -- which uses a curvature wavefront sensor driving a DM with only 188 actuators -- with a much faster (2 kHz) and far higher-order DM (64x64 actuators) driven by a Pyramid wavefront sensor.  Effectively, this two-stage SCExAO will allow the current wavefront sensor/DM loop to focus on speckle control/digging a much deeper dark hole and thus enabling a much larger planet discovery phase space.   SCExAO is also maturing near-infrared wavefront sensing capabilities to better target optically faint stars and reduce non-common path aberrations.
        \begin{figure}
      \begin{center}
  \centering
   \includegraphics[scale=0.85,clip]{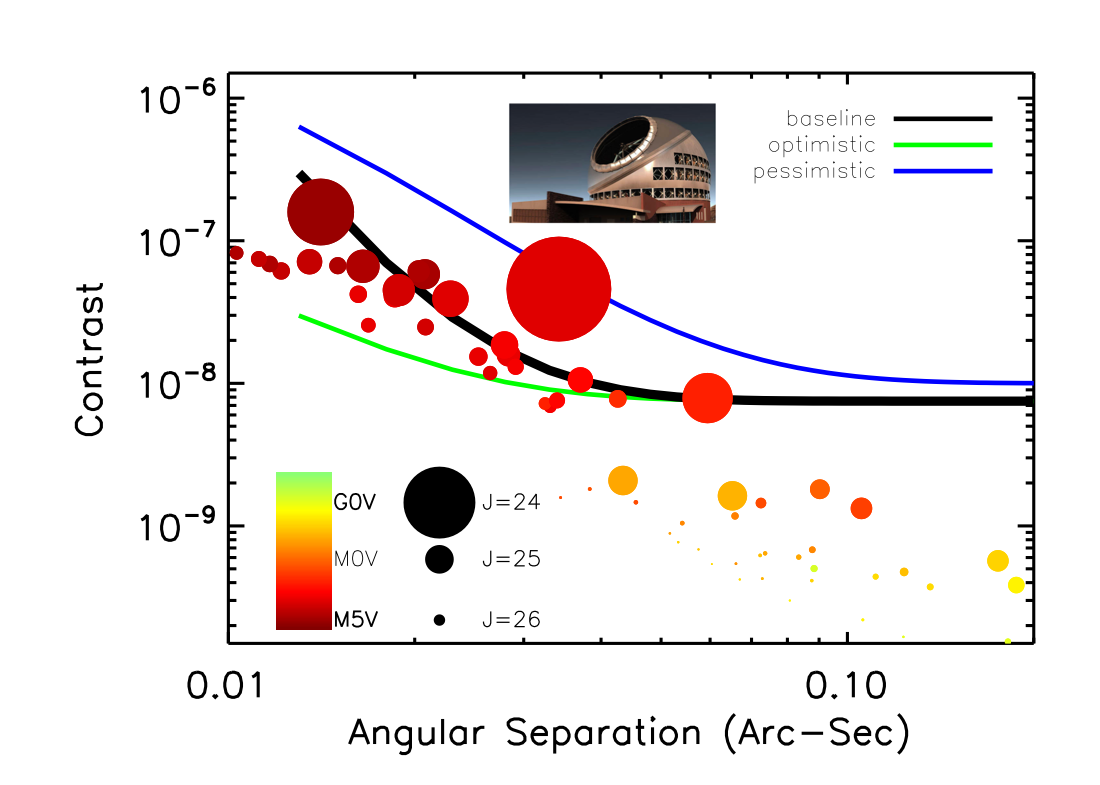}
   \end{center}
   \caption
   { \label{fig:psi} 
   Contrast curve in $J$-band for a successor instrument to SCExAO put on the TMT -- the \textit{Planetary Systems Imager} (PSI) -- using three different assumptions about its performance at small angular separations.   Systems like SCExAO help mature wavefront sensing/control methods and coronagraphy needed to achieve these performances.   Under optimistic performance scenarios ($\sim$ 10$^{-6}$ raw contrast and $\sim$ 3$\times$10$^{-8}$ contrast after post-processing at $\sim$ 1.5--2 $\lambda$/D), TMT/PSI could image rocky, Earth sized planets with an Earth-like insolation around about 20-25 nearby M stars.   
}
   \end{figure}   

\subsection{An Exoplanet Survey with SCExAO/CHARIS and the Path to Imaging Reflected-Light Planets with ELTs}

Our current plan is to carry out an exoplanet direct imaging survey using an upgraded SCExAO and the CHARIS integral field spectrograph once the system has achieved significantly deeper contrasts than demonstrated by GPI or SPHERE (e.g. 10$^{-6}$ at 0.2", 10$^{-7}$ at wide separations).   SCExAO performs substantially better in good seeing conditions.   Therefore, one option is to conduct the survey in queue mode with other programs, for instance the current InfraRed Doppler (IRD) instrument conducting a near-IR precision radial-velocity survey of the nearest low-mass stars.   

Current, blind surveys with GPI and SPHERE have resulted in a small yield of new exoplanet and low-mass brown dwarf discoveries, which limits the sample of companions with diverse properties, and thus prevents us from better mapping out the atmospheric evolution of supermassive jovian planets.    However, target selection could be modified to substantially increase yields.  For instance, targeted searches could prioritize stars showing evidence for a gravitational pull from an unseen substellar companion.   The Hipparcos-GAIA Catalog of Accelerations (HGCA) \cite{Brandt2018} provides a current master list of stars showing clear dynamical evidence for stellar to substellar companions.    Upcoming GAIA data releases will identify more weakly accelerating stars hosting lower-mass planets.   From a limited retargeting of bright, nearby HGCA stars we have already identified at least one new substellar companion using SCExAO/CHARIS.   Focusing on the sample on stars already showing evidence for companions will prevent us from deriving strong, unbiased statistics on exoplanet frequency in exchange for a much higher yield of detections to better constrain planet atmosphere evolution.
 
 Instruments like SCExAO as well as other upcoming systems like the \textit{Keck Planet Imager and Characterizer} (KPIC) and MagAO-X mature key wavefront sensing and control technology that will be implemented on ELTs to directly image rocky planets in reflected light around the nearest stars.   Habitable zone Earth-sized planets have contrasts of $\sim$ 10$^{-7}$--10$^{-8}$ but lie at angular separations of $\sim$ 0.01"--0.1" around M stars at $\sim$ 5 $pc$ or $\sim$ 1--5 $\lambda$/D for facilities like the \textit{Thirty Meter Telescope} (TMT) and \textit{Giant Magellan Telescope} (GMT)  (Figure \ref{fig:psi}).   
 
 Key contributors to the wavefront error budget and thus limiting raw contrast at small angular separations include errors due to optical path length differences (between that of the wavefront sensor and science instrument wavelengths), AO servo lag and wavefront sensor noise, and scintillation (conversion of phase errors to amplitude modulation) which are predicted to operate at the 10$^{-3}$--10$^{-4}$, 10$^{-5}$, and 10$^{-6}$ contrast levels, respectively within 3--5 $\lambda$/D \cite{Guyon2018}.   Advances being matured on SCExAO address each of these terms in the error budget.   For instance, focal-plane wavefront sensing/control being matured with SCExAO can help reduce optical path length errors; predictive control reduces the AO servo lag contribution. 
 
 Contrast curves in Figure \ref{fig:psi} depict a range of performances for the \textit{Planetary Systems Imager} -- coarsely conceived as a successor to systems like SCExAO -- on TMT depending on the system's efficacy with reducing each of these errors assuming a factor of $\sim$ 30 for speckle suppression from post-processing.  In the pessimistic case where substantial non-common path and servo lag/wavefront sensor noise contributions remain,   TMT/PSI would detect perhaps one Earth-analogue around a nearby low-mass star but many more super Earths and warm Neptunes.   For an optimistic case where TMT/PSI is able to substantially eliminate non-common path, servo lag, and wavefront sensor noise contributions, Earth-sized habitable zone planets could be detectable around nearly two-dozen M stars.   

\acknowledgments 
 
T.C. is supported by a NASA Senior Postdoctoral Fellowship and NASA/Keck grant LK-2663-948181.  We emphasize the pivotal cultural role and reverence that the summit of Maunakea has always had within the Hawaiian community.  We are most fortunate to conduct scientific observations from this mountain.   

\bibliography{report} 
\bibliographystyle{spiebib} 

\end{document}